\UseRawInputEncoding
\documentclass[%
 reprint,
superscriptaddress,
 amsmath,amssymb,
 aps,
]{revtex4-2}

\usepackage{graphicx}
\usepackage{dcolumn}
\usepackage{bm}
\usepackage{hyperref}

\begin{document}

\preprint{APS/123-QED}

\title{Controllable non-Hermitian topology in a dynamically protected cat qubit}

\author{Tian-Le Yang}
\affiliation{Fujian Key Laboratory of Quantum Information and Quantum Optics, Fuzhou University, Fuzhou 350108, China}
\affiliation{Department of Physics, Fuzhou University, Fuzhou 350108, China}
\author{Pei-Rong Han} \email{hanpeirong@lyun.edu.cn}
\affiliation{School of Physics and Mechanical and Electrical Engineering, Longyan University, Longyan 364012, China}
\author{Zhen-Biao Yang} \email{zbyang@fzu.edu.cn}
\affiliation{Fujian Key Laboratory of Quantum Information and Quantum Optics, Fuzhou University, Fuzhou 350108, China}
\affiliation{Department of Physics, Fuzhou University, Fuzhou 350108, China}
\affiliation{Hefei National Laboratory, Hefei 230088, China}
\author{Shi-Biao Zheng} \email{t96034@fzu.edu.cn}
\affiliation{Fujian Key Laboratory of Quantum Information and Quantum Optics, Fuzhou University, Fuzhou 350108, China}
\affiliation{Department of Physics, Fuzhou University, Fuzhou 350108, China}
\affiliation{Hefei National Laboratory, Hefei 230088, China}

\date{\today}

\begin{abstract}
Dissipatively stabilized cat qubits are promising for fault-tolerant quantum information processing, yet their non-Hermitian (NH) spectral topology remains largely unexplored. We uncover rich Liouvillian exceptional structures in a cat-qubit mode stabilized by two-photon drive (TPD) and engineered two-photon loss, in the presence of single-photon drive (SPD) and single-photon loss. In the parameter space spanned by SPD strength and detuning, we identify both second- and third-order Liouvillian exceptional points (LEP2s and LEP3s). Remarkably, we show that the phase $\theta$ of TPD provides coherent control over these exceptional points: the LEP3 diverges and vanishes at $\theta=\pi/2$, while remaining stable and tunable elsewhere. We introduce a topological invariant based on the winding number of a resultant vector, which robustly identifies LEP3s with unit topological charge. Full master-equation simulations confirm that the system dynamics remains confined to the logical subspace with near-unity fidelity. Our results bridge dissipative stabilization, phase-coherent control, and NH topology, demonstrating controllable higher-order LEPs in open quantum systems.
\end{abstract}

\maketitle


\section{\label{sec:Introduction}Introduction}

Non-Hermitian (NH) physics has emerged as a cornerstone for understanding open quantum systems, with exceptional points (EPs)---degeneracies where eigenvalues and eigenvectors coalesce---playing a central role~\cite{Dembowski2001,Zhou2018,Cerjan2019,Zhen2015,Wang2021,Zhang2023,Tang2021,Su2021,Zhang2021,Cao2023,Wu2023,Tang2020,Tang2023,Han2024,Zhang2025}.
In quantum settings, EPs manifest in two distinct operational contexts:
(i) Hamiltonian EPs (HEPs), which arise in effective NH Hamiltonians describing measurement-conditioned or post-selected evolution, where quantum jumps are excluded by conditioning~\cite{Han2024,Naghiloo2019,Abbasi2022,Han2023,ZhangGuo2025,Yan2025};
(ii) Liouvillian EPs (LEPs), which occur in the spectrum of the full Liouvillian superoperator governing unconditional open-system dynamics, inherently accounting for all dissipative quantum trajectories~\cite{Minganti2019,Arkhipov2020,Minganti2020,Arkhipov2020a,Khandelwal2021,Sun2024,Chen2021,Chen2022a,Abo2024,Zhang2022,Bu2023}.
While LEPs have been extensively explored in discrete-variable qubits~\cite{Minganti2019,Arkhipov2020,Minganti2020,Arkhipov2020a,Khandelwal2021,Sun2024,Chen2021,Chen2022,Abo2024,Zhang2022,Bu2023}, their investigation in continuous-variable (CV) systems remains largely in its early stages~\cite{Han2025}.

A promising platform for CV quantum information is the cat qubit, encoded in superpositions of coherent states $|\mathcal{C}^\pm_\alpha\rangle = \mathcal{N}^\pm_\alpha \left( |\alpha\rangle \pm |{-}\alpha\rangle \right)$, where $\alpha$ is the complex amplitude (or ``cat size'') and $\mathcal{N}^\pm_\alpha$ normalization constants.
This encoding features a noise bias, with bit-flip errors exponentially suppressed as $\alpha$ increases~\cite{Grimm2020,Puri2020,Reglade2024}. Cat qubits can be stabilized via distinct mechanisms: intrinsic Kerr nonlinearity with two-photon drive (Kerr-cat scheme)~\cite{Vlastakis2013,Goto2016,Puri2017,Puri2019,Grimm2020}, or engineered dissipation that autonomously confines the system to the logical subspace (dissipative cat scheme)~\cite{Mirrahimi2014,Leghtas2015,Albert2016,Lescanne2020,Berdou2023,Reglade2024}. Crucially, both approaches yield an effective two-dimensional logical manifold spanned by $|\mathcal{C}^\pm_\alpha\rangle$, enabling the Liouvillian dynamics to be projected onto a common $4\times4$ matrix structure in the vectorized density-operator space. 
Recent works identified second- and third-order Liouvillian exceptional points (LEP2s and LEP3s) in Kerr‑based cat qubits~\cite{Han2025,Han2026}. Here we show that dissipatively stabilized cat qubits host LEP2s and LEP3s as well, and moreover, that these exceptional structures can be controlled by the phase of the two‑photon drive---a control parameter beyond drive strength and detuning.

We investigate the NH spectral topology of a cat-qubit system stabilized by two-photon drive (TPD) and engineered two-photon loss (ETPL), and subject to single-photon drive (SPD) and single-photon loss (SPL).
The interplay among SPD, detung, SPL, and the TPD phase gives rise to controllable Liouvillian exceptional structures, including LEP2s and LEP3s.
Their positions and related topology can be tuned via a classical phase parameter. By introducing a winding number based on a resultant vector, we construct a robust topological invariant that characterizes LEP3s in open quantum systems. Full master-equation simulations rigorously validate the logical subspace approximation, demonstrating high-fidelity dynamical confinement. Our results establish the ETPL cat qubit as a versatile platform for exploring higher-order exceptional physics and intrinsic topology.

\begin{figure}
	\includegraphics[width =0.48\textwidth]{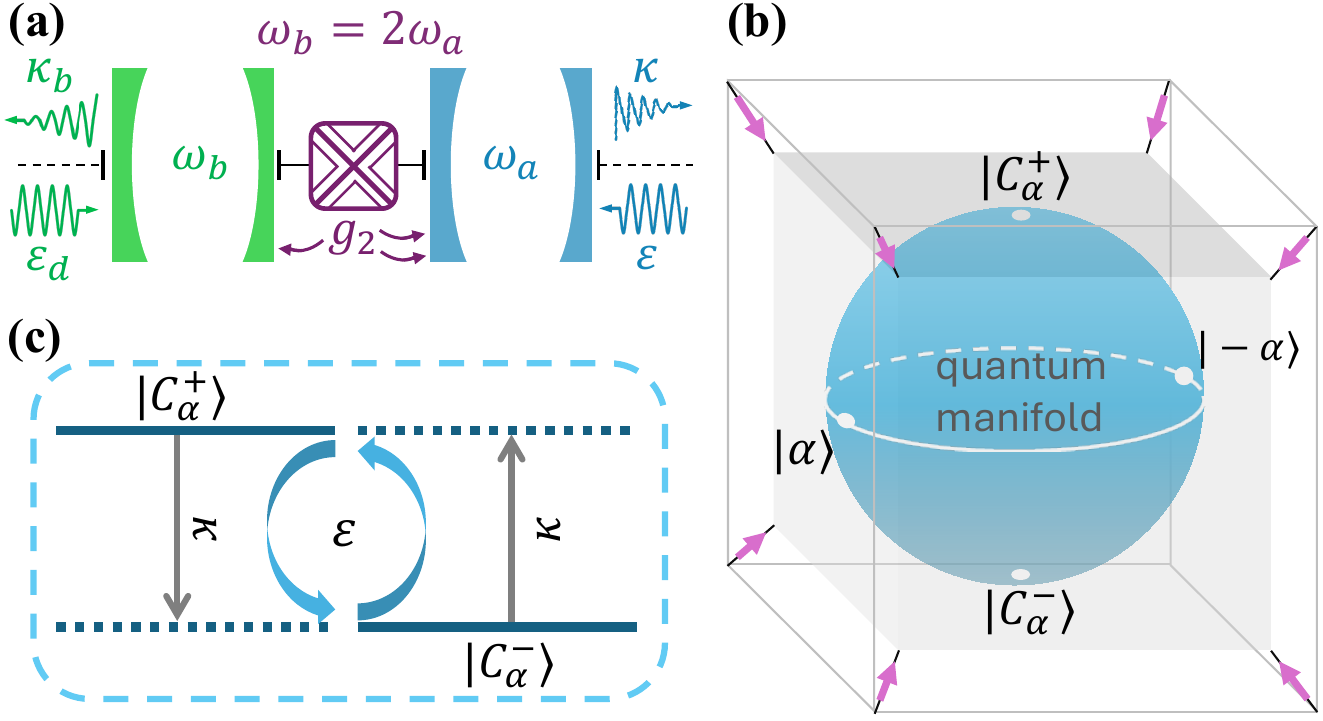}
	\caption{\label{Fig1} (a) Implementation of the target dynamics via a three-wave mixing coupler $\omega_b = 2\omega_a$ that nonlinearly couples the memory mode (blue cavity) to an SPD-driven lossy buffer mode (green cavity) with strength $g_{2}$.
    Adiabatic elimination of the buffer yields ETPL and TPD on the memory mode, while an additional independent SPD applied directly to the memory enables coherent control of the Liouvillian spectrum.
    (b) Dissipative confinement of the oscillator state to the 2D manifold spanned by the coherent states $|\pm\alpha\rangle$ (blue sphere), which contains the even and odd Schr\"odinger cat states $|\mathcal{C}^\pm_\alpha\rangle$ as the logical basis. Stabilization via TPD and ETPL drives all states toward this subspace without inducing coherent rotations, as indicated by the purple arrows.
    (c) Dynamics within the cat-qubit subspace: a weak SPD (strength $\varepsilon$) drives coherent Rabi oscillations between $|\mathcal{C}^{+}_{\alpha}\rangle$ and $|\mathcal{C}^{-}_{\alpha}\rangle$, while SPL (rate $\kappa$) induces incoherent bit-flip errors.
	}
\end{figure}


\section{\label{sec:Model}Model and subspace encoding}

We consider a bosonic mode stabilized by TPD (strength $\varepsilon_2 = |\varepsilon_2| e^{-i\theta}$) and ETPL (rate \(\kappa_2\)), and subject to SPD (strength $\varepsilon$, taken real) and SPL (rate \(\kappa\)). In the frame rotating at the SPD frequency, the dynamics is governed by the Lindblad master equation ($\hbar=1$ is set):
\begin{equation}
\dot{\rho} = -i[H, \rho] + \kappa \, \mathcal{D}[a] \rho + \kappa_2 \, \mathcal{D}[a^2] \rho,
\label{eq:master}
\end{equation}
with Hamiltonian
\begin{equation}
H = \Delta a^\dagger a + (\varepsilon_2 a^{\dagger 2} + \varepsilon_2^* a^2) + \varepsilon(a^\dagger + a),
\label{eq:H_new}
\end{equation}
and $\mathcal{D}[\mathcal{O}]\rho = \mathcal{O}\rho \mathcal{O}^\dagger - \frac{1}{2}\{\mathcal{O}^\dagger \mathcal{O}, \rho\}$ for $\mathcal{O}=a,\,a^{2}$, where $a$ ($a^{\dagger}$) is the annihilation (creation) operator for the bosonic mode. Here $\Delta$ is the detuning between the SPD and the bosonic mode. This master equation can be implemented in superconducting circuits via three-wave~\cite{Marquet2024,Marquet2024a} or four-wave~\cite{Leghtas2015,Touzard2018,Lescanne2020,Berdou2023,Reglade2024} mixing.
Taking the three-wave mixing process as an example, Fig.~\ref{Fig1}(a) illustrates the setup: an SPD is applied to the memory mode in an autoparametric superconducting circuit. The memory mode is coupled to a lossy buffer mode (annihilation operator $ b $, frequency $ \omega_b $, SPL rate $ \kappa_b $), whose frequency is set to twice that of the memory mode, giving the nonlinear coupling $ H_{\text{a,b}} = g_2 a^2 b^\dagger + \text{H.c.} $~\cite{Marquet2024,Marquet2024a}.
In the regime $\kappa_b \geq 8 g_2 |\alpha|$, the buffer mode can be adiabatically eliminated, yielding an effective TPD with strength $\varepsilon_{2}=-2ig_{2}\varepsilon_{d}/\kappa_{b}$ and an effective ETPL with rate $\kappa_2 = 4g_2^2 / \kappa_b$ (see Ref.~\cite{Leghtas2015} for the derivation).

The combination of ETPL and TPD stabilizes Schr\"{o}dinger cat states. In the absence of all other terms ($\Delta = \varepsilon = \kappa = 0$), photon-number parity is conserved: TPD injects photons only in pairs, while ETPL removes them in pairs. An initial even-parity state, such as the vacuum $|0\rangle\langle0|$, converges to the even cat state $|\mathcal{C}^{+}_\alpha\rangle$ with amplitude $\alpha = \sqrt{-2i \varepsilon_2/\kappa_2 } = \sqrt{2|\varepsilon_2|/\kappa_2}\, e^{i\varphi_{\alpha}}$, where $\varphi_{\alpha} = 3\pi/4 - \theta/2$. Similarly, an odd-parity initial state (e.g., $|1\rangle\langle1|$) converges to the odd cat state $|\mathcal{C}^{-}_\alpha\rangle$~\cite{Gilles1994,Mirrahimi2014,Leghtas2015}. These two states define the logical cat-qubit subspace $\{|\mathcal{C}^{+}_\alpha\rangle, |\mathcal{C}^{-}_\alpha\rangle\}$, illustrated in Fig.~\ref{Fig1}(b).
For any initial state, the system converges exponentially to the two-dimensional (2D) manifold spanned by $\{|\mathcal{C}^{+}_\alpha\rangle, |\mathcal{C}^{-}_\alpha\rangle\}$ (blue Bloch sphere in Fig.~\ref{Fig1}(b)) with confinement rate $\kappa_{\text{conf}} = 4|\alpha|^2 \kappa_2$ in the limit $\kappa_2 \gg \kappa$.

A weak SPD ($\varepsilon \ll \kappa_2$) induces coherent Rabi oscillations between $|\mathcal{C}^{+}_\alpha\rangle$ and $|\mathcal{C}^{-}_\alpha\rangle$, realizing controllable logical rotations (e.g., an $X$ rotation) rather than errors [Fig.~\ref{Fig1}(c)].
In contrast, natural SPL at rate $\kappa$ is the primary source of non-Hermiticity and causes bit-flip errors: the projected annihilation operator behaves as
$a = \alpha \left( p\, |\mathcal{C}^+_\alpha\rangle\langle \mathcal{C}^-_\alpha| + p^{-1}\, |\mathcal{C}^-_\alpha\rangle\langle \mathcal{C}^+_\alpha| \right)$, with $p=\mathcal{N}_{\alpha}^+/\mathcal{N}_{\alpha}^-$, so that a single-photon jump flips the logical state [Fig.~\ref{Fig1}(c)].
A detuning $\Delta$ lifts the degeneracy of the cat states, generating an effective $\sigma_z$ Hamiltonian that drives coherent rotations around the logical $Z$-axis.

The interplay between coherent SPD, dissipative SPL, and dispersive detuning within the ETPL-stabilized subspace gives rise to rich Liouvillian spectral structures, including second-order exceptional lines (LEP2s) and third-order exceptional points (LEP3s). We characterize their topology via the winding number of a resultant vector~\cite{Delplace2021}.

\section{Liouvillian exceptional structures}

To analyze the Liouvillian spectrum and locate LEPs, we adopt the vectorized representation $\rho \mapsto \boldsymbol{V} = \mathrm{vec}(\rho)$, under which the master equation~\eqref{eq:master} becomes a linear ordinary differential equation $\dot{\boldsymbol{V}}(t) = \mathcal{L}_{\text{matrix}} \boldsymbol{V}(t)$,
where the Liouvillian matrix is
\begin{eqnarray}
\mathcal{L}_{\text{matrix}} &=& -i\left(H \otimes \mathbb{I} - \mathbb{I} \otimes H^\top\right)\nonumber+\sum_{\mathcal{O}=a,\,a^{2}}\Big[\Gamma_{\mathcal{O}} \otimes \Gamma_{\mathcal{O}}^{*}\\
& &- \frac{1}{2}\,\Gamma_{\mathcal{O}}^\dagger \Gamma_{\mathcal{O}} \otimes \mathbb{I}
- \frac{1}{2}\,\mathbb{I} \otimes (\Gamma_{\mathcal{O}}^\top \Gamma_{\mathcal{O}}^{*})\Big],
\end{eqnarray}
with $\Gamma_a = \sqrt{\kappa}\,a$, $\Gamma_{a^2} = \sqrt{\kappa_2}\,{a^2}$.

Projecting onto the logical subspace $\{ |\mathcal{C}^+_\alpha\rangle, |\mathcal{C}^-_\alpha\rangle \}$ and using the identities
\begin{equation}
a |\mathcal{C}^\pm_\alpha\rangle = \alpha p^{\mp 1} |\mathcal{C}^\mp_\alpha\rangle,\, a^\dagger |\mathcal{C}^\pm_\alpha\rangle = \alpha^{*} p^{\pm 1} |\mathcal{C}^\mp_\alpha\rangle,
\end{equation}
the projected Liouvillian reduces to a $4\times4$ matrix:
\begin{widetext}
\begin{equation}
\mathcal{L}_{\text{matrix}} =
\begin{pmatrix}
-\kappa|\alpha|^{2}p^{2} & 
i\varepsilon(\alpha p + \alpha^{*}p^{-1}) & 
-i\varepsilon(\alpha^{*}p + \alpha p^{-1}) & 
\kappa|\alpha|^{2}p^{-2} \\
i\varepsilon(\alpha^{*}p + \alpha p^{-1}) & 
i\Delta|\alpha|^{2}p_{2}^{-} - \dfrac{\kappa}{2}|\alpha|^{2}p_{2}^{+} & 
\kappa|\alpha|^{2} & 
-i\varepsilon(\alpha^{*}p + \alpha p^{-1}) \\
-i\varepsilon(\alpha p + \alpha^{*}p^{-1}) & 
\kappa|\alpha|^{2} & 
-i\Delta|\alpha|^{2}p_{2}^{-} - \dfrac{\kappa}{2}|\alpha|^{2}p_{2}^{+} & 
i\varepsilon(\alpha p + \alpha^{*}p^{-1}) \\
\kappa|\alpha|^{2}p^{2} & 
-i\varepsilon(\alpha p + \alpha^{*}p^{-1}) & 
i\varepsilon(\alpha^{*}p + \alpha p^{-1}) & 
-\kappa|\alpha|^{2}p^{-2}
\end{pmatrix},
\end{equation}
\end{widetext}
where $p^{\pm}_j = p^{-j} \pm p^{j}$ are introduced for compactness.
The non-Hermiticity arises from SPL; ETPL and the TPD phase set the cat amplitude $\alpha$ and thereby control the Liouvillian exceptional structures.

Assuming diagonalizability, the time evolution is
\begin{equation}
\boldsymbol{V}(t) = \sum_{k=1}^4 c_k e^{E_k t} \boldsymbol{V}_k,
\label{eq:V(t)}
\end{equation}with $E_k$ being the eigenvalues of $\mathcal{L}_{\text{matrix}}$, $\boldsymbol{V}_k$ the corresponding eigenvectors, and $c_k$ the expansion coefficients determined by the initial condition.
The characteristic polynomial yields the four closed-form eigenvalues:
\begin{align}
E_1 &= 0, \\
E_2 &= -\frac{2}{3}\,\kappa |\alpha|^2 p^{+}_2 + (\eta_+ + \eta_-), \\
E_3 &= -\frac{2}{3}\,\kappa |\alpha|^2 p^{+}_2 + \left(e^{i 2\pi/3} \eta_+ + e^{-i 2\pi/3} \eta_-\right), \\
E_4 &= -\frac{2}{3}\,\kappa |\alpha|^2 p^{+}_2 + \left(e^{-i 2\pi/3} \eta_+ + e^{i 2\pi/3} \eta_-\right),
\end{align}
where $\eta_\pm=(q \pm \sqrt{q^2 + m^3})^{1/3}$,
and $q$ and $m$ are given by
\begin{align}
q &= \frac{|\alpha|^{2} \kappa}{216} \Big[ 
    -|\alpha|^{4} (36\Delta^{2} + \kappa^{2}) p^{+}_{6} 
    + 72 |\alpha|^{2} \varepsilon^{2} p^{+}_{4} \nonumber \\
&\quad + \big(  
    36 |\alpha|^{4} \Delta^{2} 
    + 33 |\alpha|^{4} \kappa^{2} 
    -576 \varepsilon^{2} |\alpha|^{2} \sin(\theta)\big) p^{+}_{2} \nonumber \\
&\quad    + 1008 |\alpha|^{2} \varepsilon^{2} 
\Big],
\label{q}
\end{align}and
\begin{align}
m &= \frac{1}{36} \Big[ 
    |\alpha|^{4} (12\Delta^{2} - \kappa^{2}) p^{+}_{4} 
    + 48 |\alpha|^{2} \varepsilon^{2} p^{+}_{2} \nonumber \\
&\quad - 96 \varepsilon^{2} |\alpha|^{2} \sin(\theta)  
    - |\alpha|^{4} (24\Delta^{2} + 14\kappa^{2}) 
\Big],
\label{m}
\end{align}respectively.
The eigenvalue $E_1 = 0$ corresponds to the steady state; $E_2$ is purely real; $E_3$ and $E_4$ form a complex-conjugate pair. An LEP2 occurs when $E_3 = E_4$; an LEP3 appears when $E_2 = E_3 = E_4$.
For $\varepsilon = 0$, an LEP2 exists at $|\Delta| = \kappa / p_2^{-}$, independent of $\theta$. However, for $\varepsilon \neq 0$, SPD induces a topological transition: the original LEP2 splits into two LEP2s of the same order whose positions become $\theta$-dependent via $\alpha$.
The conditions for LEP3s are
\begin{align}
\varepsilon^{2}_{|\alpha|,\,\theta} 
&= \frac{(1 + p^{4})^{3} \, \kappa^{2} \, |\alpha|^{2}}{54 \, p^{2} \, \mathcal{D}_{\theta}}, \label{LEP3_var_theta} \\ 
\Delta^{2}_{|\alpha|,\,\theta} 
&= \frac{\kappa^{2}}{108 \, (p^{4} - 1)^{2} \, \mathcal{D}_{\theta}} 
   \Big[ (1 + 148 p^{4} + 726 p^{8} \nonumber \\
&\quad + 148 p^{12} + p^{16}) - 4 p^{2} (1 + p^{4}) \nonumber \\
&\quad \times(5 + 118 p^{4} + 5 p^{8}) \sin(\theta) \Big], \label{LEP3_Delta_theta}
\end{align}
with $\mathcal{D}_{\theta} = (1+p^4)^2+4p^4-4p^2(1+p^4)\sin(\theta)$.
For $\theta=\pi/2$, $\alpha$ is purely imaginary and $\mathcal{D}_{\theta}\rightarrow 0$ as $|\alpha|\rightarrow \infty$ ($p\rightarrow1$), causing $|\varepsilon_{|\alpha|,\,\theta}|\rightarrow \infty$: the LEP3 disappears. 
For $\theta \neq \pi/2$, the LEP3 persists and its location varies continuously with $\theta$, demonstrating controllable exceptional structures. For $\theta = \theta_0 = 3\pi/2$ (real $\alpha$), the expressions simplify to
\begin{align}
|\varepsilon_{|\alpha|,\,\theta_{0}}| &= \frac{\sqrt{6}\,\kappa\,|\alpha|\,(p^{4}+1)^{3/2}}{18\,p\,(p^{2}+1)^{2}}, \\
|\Delta_{|\alpha|,\,\theta_{0}}| &= \frac{\sqrt{3}\,\kappa\,(p^{4}+6p^{2}+1)^{3/2}}{18\,(p^{2}-1)\,(p^{2}+1)^{2}}.
\end{align}
For other values of $\theta$ (excluding $\pi/2$, $3\pi/2$), $\alpha$ is complex and both $|\varepsilon_{|\alpha|,\,\theta}|$ and $|\Delta_{|\alpha|,\,\theta}|$ exhibit periodic modulation with $\theta$, reflecting the rotational symmetry of the cat-state orientation in phase space.

\begin{figure*}
	\includegraphics[width =\textwidth]{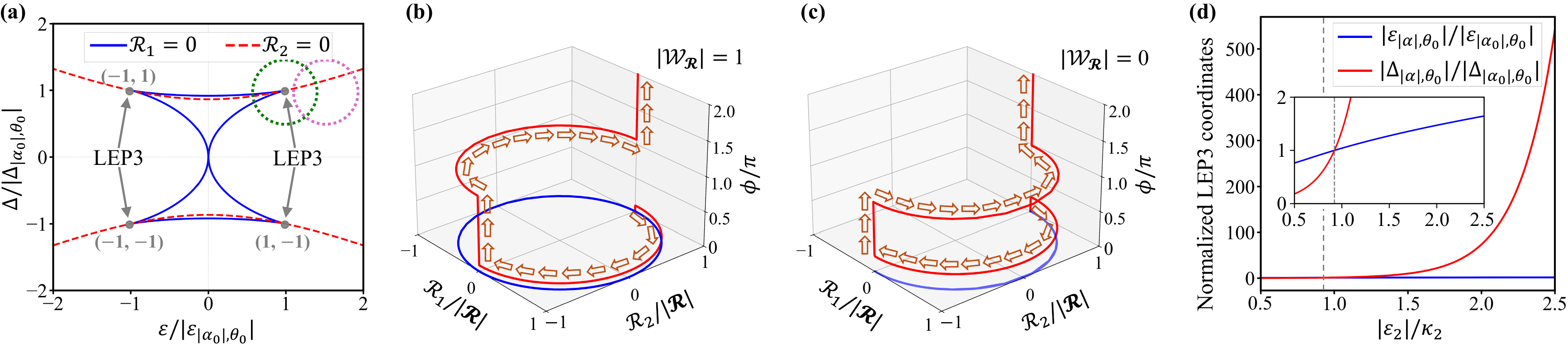}
	\caption{\label{Fig2} Liouvillian exceptional structures in the ETPL cat-qubit model.
(a) Contour lines of $\mathcal{R}_1 = 0$ and $\mathcal{R}_2 = 0$ in the plane of $(\varepsilon/|\varepsilon_{|\alpha_{0}|,\,\theta_{0}}|,\, \Delta/|\Delta_{|\alpha_{0}|,\,\theta_{0}}|)$. 
(b)--(c) Winding number of the resultant vector $\boldsymbol{\mathcal{R}} = (\mathcal{R}_1,\,\mathcal{R}_2)$.  
(b) Trajectory of the normalized resultant vector $\mathcal{R}_{\text{N}} \equiv (\mathcal{R}_1 + i\mathcal{R}_2)/|\boldsymbol{\mathcal{R}}|$ as the system parameters are varied along the loop $\mathcal{C}_\phi$. The loop is defined by $\Delta(\phi) = 0.4\,\Delta_{|\alpha_{0}|,\,\theta_0} \sin(\phi) + \Delta_{|\alpha_{0}|,\,\theta_0}$ and $\varepsilon(\phi) = 0.4\,\varepsilon_{|\alpha_{0}|,\,\theta_0} \cos(\phi) + \varepsilon_{|\alpha_{0}|,\,\theta_0}$, corresponding to the green dotted circle in (a). The $x$-, $y$-, and $z$-axes represent $\mathcal{R}_1/|\boldsymbol{\mathcal{R}}|$, $\mathcal{R}_2/|\boldsymbol{\mathcal{R}}|$, and $\phi/\pi$, respectively. The red curve traces the evolution of $\mathcal{R}_1$ and $\mathcal{R}_2$ with $\phi$; its projection onto the rescaled $\mathcal{R}_1$-$\mathcal{R}_2$ plane (solid blue curve) forms a closed loop, indicating a topological winding number $|\mathcal{W}_{\boldsymbol{\mathcal{R}}}| = 1$. 
(c) Same as (b), but with $\varepsilon(\phi) = 0.4\,\varepsilon_{|\alpha_{0}|,\,\theta_0} \cos(\phi) + 1.5\,\varepsilon_{|\alpha_{0}|,\,\theta_0}$, corresponding to the purple dotted circle in (a). In this case, the projected trajectory does not enclose the origin and does not form a closed loop, yielding a winding number $|\mathcal{W}_{\boldsymbol{\mathcal{R}}}| = 0$. 
(d) Normalized LEP3 coordinates ($|\varepsilon_{|\alpha|,\,\theta_{0}}|/|\varepsilon_{|\alpha_{0}|,\,\theta_{0}}|$, $|\Delta_{|\alpha|,\,\theta_{0}}|/|\Delta_{|\alpha_{0}|,\,\theta_{0}}|$) as functions of $|\varepsilon_{2}|/\kappa_{2}$.
The gray dashed line marks $|\varepsilon_{2}|/\kappa_{2}\approx0.93$, corresponding to the reference value $|\alpha_{0}|\approx1.36$.
In all panels, we set $\kappa/\kappa_{2}\approx6.48\times10^{-3}$, $\theta_{0}=3\pi/2$, and $|\alpha_{0}|\approx1.36$.
}
\end{figure*}

\begin{figure*}
	\includegraphics[width =\textwidth]{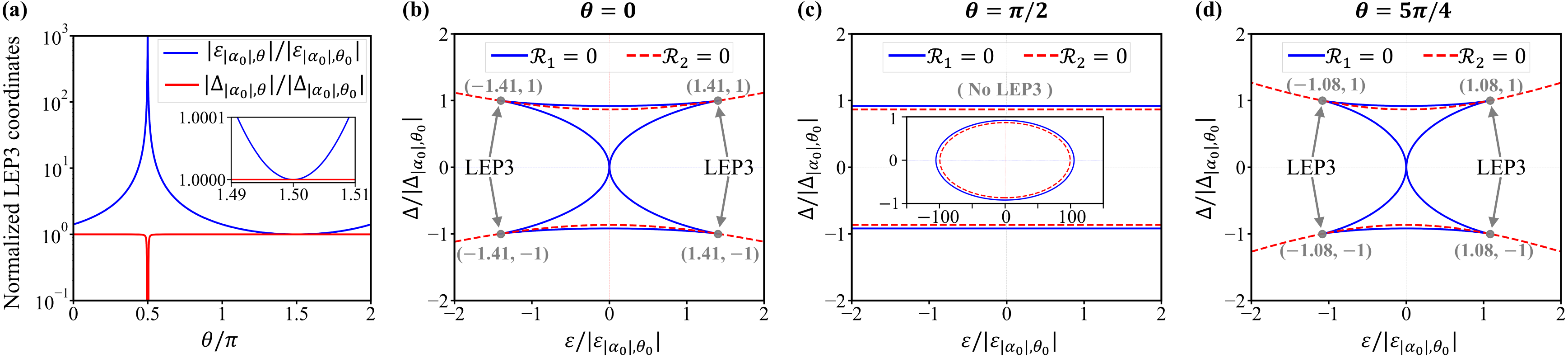}
	\caption{\label{Fig3} Phase-controlled Liouvillian exceptional structures.
    (a) Normalized LEP3 coordinates ($|\varepsilon_{|\alpha_{0}|,\,\theta}|/|\varepsilon_{|\alpha_{0}|,\,\theta_{0}}|$, $|\Delta_{|\alpha_{0}|,\,\theta}|/|\Delta_{|\alpha_{0}|,\,\theta_{0}}|$) as functions of $\theta/\pi$. The inset shows the magnified effect near $\theta = 1.5 \pi$.
    (b)--(d) Phase-controlled LEP2 lines ($\mathcal{R}_{1}=0$) and LEP3s ($\mathcal{R}_{1}=\mathcal{R}_{2}=0$).
    The zero-contours $\mathcal{R}_{1}=0$ and $\mathcal{R}_{2}=0$ are plotted as functions of $\varepsilon/|\varepsilon_{|\alpha_{0}|,\theta_{0}}|$ for $\theta=0$~(b), $\theta=\pi/2$~(c), and $\theta=5\pi/4$~(d).
    In all panels, $\kappa/\kappa_{2}\approx6.48\times10^{-3}$, $\theta_{0}=3\pi/2$, and $|\alpha_{0}|\approx1.36$.
	}
\end{figure*}

To characterize the topological nature of the exceptional structures, we use the winding number of the resultant vector $\boldsymbol{\mathcal{R}} = (\mathcal{R}_1, \mathcal{R}_2)$ ~\cite{Delplace2021}:
\begin{equation}
\mathcal{W}_{\boldsymbol{\mathcal{R}}}=\frac{1}{2\pi}\oint_{\mathcal{C}_{\phi}}\frac{1}{\lVert \boldsymbol{\mathcal{R}} \rVert^{2}}\left(\mathcal{R}_{1}\frac{\partial\mathcal{R}_{2}}{\partial\phi}-\mathcal{R}_{2}\frac{\partial\mathcal{R}_{1}}{\partial\phi}\right)d\phi,
\end{equation}
where $\mathcal{C}_{\phi}$ is a closed contour in $(\varepsilon,\,\Delta)$ space dependent on $\phi$.
The characteristic polynomial is 
$P(E)=\prod_{i=1}^{4}(E-E_{i})$.
Factoring out the steady-state eigenvalue, the cubic $\tilde{P}(E)=(E-E_{2})(E-E_{3})(E-E_{4})$ yields
\begin{equation}
\mathcal{R}_{1}=-(E_{2}-E_{3})^{2}(E_{2}-E_{4})^{2}(E_{3}-E_{4})^{2},
\end{equation}
\begin{align}
\mathcal{R}_{2}&=-8(E_{2}+E_{3}-2E_{4})(E_{2}+E_{4}-2E_3)\nonumber \\
 &\quad\times(E_{3}+E_{4}-2E_{2}).
\end{align}
$\mathcal{R}_1 = 0$ signals an LEP2 (any two eigenvalues coincide); and $\mathcal{R}_1 = \mathcal{R}_2 = 0$ signals an LEP3 (all three coincide).
As shown in Ref.~\cite{Delplace2021}, $|\mathcal{W}_{\boldsymbol{\mathcal{R}}}| = 1$ around a closed contour indicates a single LEP3 inside. The zero contours of $\mathcal{R}_1$ and $\mathcal{R}_2$ in the $(\varepsilon,\,\Delta)$ plane, plotted in Figs.~\ref{Fig2} and \ref{Fig3}, delineate LEP2 lines and pinpoint LEP3s at their intersections.

\subsection{Liouvillian exceptional structures in ETPL cat-qubit model}

We first show that LEPs with nontrivial topological structure can emerge in the ETPL cat-qubit model. 
Figure~\ref{Fig2}(a) presents the zero-contours of $\mathcal{R}_1$ and $\mathcal{R}_2$ in the parameter space spanned by the dimensionless variables $\varepsilon / |\varepsilon_{|\alpha_0|,\,\theta_0}|$ and $\Delta / |\Delta_{|\alpha_0|,\,\theta_0}|$, with $|\alpha_0| \approx 1.36$ and $\theta_0 = 3\pi/2$. As the simplest case, we fix the TPD phase to $\theta = \theta_0$, which renders the cat amplitude $\alpha$ real. The more general case with complex $\alpha$ ($\theta \neq \theta_0$) is discussed in the following section.
To ensure broad applicability of our findings, we formulate the dynamics using dimensionless parameters normalized by the ETPL rate $\kappa_2$. 
For the analysis, we adopt representative parameters inspired by state-of-the-art circuit QED experiments~\cite{Marquet2024,Marquet2024a}: 
a normalized SPL rate $\kappa/\kappa_{2} \approx 6.48\times10^{-3}$ and a normalized TPD strength $|\varepsilon_{2}|/\kappa_{2} \approx 0.93$. 
These ratios yield a steady-state coherent-state amplitude of $|\alpha_{0}| \approx 1.36$, which defines the logical subspace for the cat qubit. 
Although specific numerical values are used for illustration, the Liouvillian exceptional structures and topological features discussed below depend primarily on these dimensionless ratios rather than absolute frequency scales.
In the absence of SPD ($\varepsilon = 0$), an LEP2 exists at $|\Delta| = \kappa / p_2^{-}$. This unperturbed LEP2 is independent of the TPD phase $\theta$, because the $\theta$-dependent terms in Eqs.~\eqref{q} and \eqref{m} vanish when $\varepsilon = 0$. When SPD is turned on ($\varepsilon \neq 0$), a topological transition occurs: the original LEP2 at $|\Delta| = \kappa / p_2^{-}$ splits into two distinct LEP2s.
The LEP3s appear at $(\varepsilon/|\varepsilon_{|\alpha_{0}|,\,\theta_{0}}|,\,\Delta/|\Delta_{|\alpha_{0}|,\,\theta_{0}}|)=(\pm1,\,\pm1)$, corresponding to the intersection points of the $\mathcal{R}_1 = 0$ and $\mathcal{R}_2 = 0$ contours.

To illustrate the topological nature of the LEP3, we plot in Figs.~\ref{Fig2}(b) and \ref{Fig2}(c) the trajectory of the normalized resultant vector $\mathcal{R}_{\text{N}} \equiv (\mathcal{R}_1 + i\mathcal{R}_2)/|\boldsymbol{\mathcal{R}}|$ as the system parameters are varied along a closed loop $\mathcal{C}_\phi$, with the TPD phase fixed at $\theta_0 = 3\pi/2$ and $|\alpha_0| \approx 1.36$.
The loop $\mathcal{C}_{\phi}$ in Fig.~\ref{Fig2}(b) is defined by $\Delta(\phi) = 0.4\,\Delta_{|\alpha_{0}|,\,\theta_0} \sin(\phi) + \Delta_{|\alpha_{0}|,\,\theta_0}$ and $\varepsilon(\phi) = 0.4\,\varepsilon_{|\alpha_{0}|,\,\theta_0} \cos(\phi) + \varepsilon_{|\alpha_{0}|,\,\theta_0}$, corresponding to the green dotted circle in Fig.~\ref{Fig2}(a).
The trajectory of $\mathcal{R}_{\mathrm{N}}$ winds once around the origin, yielding a winding number $\mathcal{W}_{\boldsymbol{\mathcal{R}}} = 1$, confirming that the enclosed region contains an LEP3.
In contrast, the loop in Fig.~\ref{Fig2}(c) uses the same $\Delta(\phi)$ but shifts the SPD strength to $\varepsilon(\phi) = 0.4\,\varepsilon_{|\alpha_{0}|,\,\theta_0} \cos(\phi) + 1.5\,\varepsilon_{|\alpha_{0}|,\,\theta_0}$, corresponding to the purple dotted circle in Fig.~\ref{Fig2}(a). Here, the trajectory does not enclose the origin, giving $\mathcal{W}_{\boldsymbol{\mathcal{R}}} = 0$, which indicates the absence of an LEP3 within the loop.

To examine how the ratio $|\varepsilon_2|/\kappa_2$---which sets the cat size $|\alpha| = \sqrt{2|\varepsilon_2|/\kappa_2}$---affects the coordinates of LEP3s, we plot in Fig.~\ref{Fig2}(d) the normalized LEP3 coordinates ($|\varepsilon_{|\alpha|, \theta_0}|/|\varepsilon_{|\alpha_0|, \theta_0}|$, $|\Delta_{|\alpha|, \theta_0}|/|\Delta_{|\alpha_0|, \theta_0}|$) as functions of $|\varepsilon_2|/\kappa_2$, with $\theta_0 = 3\pi/2$ and a reference value $|\alpha_0| \approx 1.36$ (i.e., $|\varepsilon_2|/\kappa_2 \approx 0.93$, marked by the gray dashed line).
As $|\varepsilon_2|/\kappa_2$ increases, the normalized SPD strength $|\varepsilon_{|\alpha|,\,\theta_0}|/|\varepsilon_{|\alpha_0|,\,\theta_0}|$ grows approximately linearly, whereas the normalized detuning $|\Delta_{|\alpha|,\, \theta_0}|/|\Delta_{|\alpha_0|,\,\theta_0}|$ increases more rapidly, exhibiting near-exponential growth. 
For $|\varepsilon_2|/\kappa_2 \lesssim 0.93$, both normalized coordinates lie below unity, indicating that the LEP3 resides closer to the origin in the $(\varepsilon,\,\Delta)$ plane than the reference point. Conversely, for $|\varepsilon_2|/\kappa_2 \gtrsim 0.93$, the LEP3 moves away from the reference point, with both coordinates exceeding unity and increasing rapidly as the cat size increases.

These results confirm that both LEP2 and LEP3 can be realized in the ETPL cat-qubit model. Moreover, the LEP3 exhibits the same topological signature---a unit winding number---as the third-order Hamiltonian exceptional point (HEP3) reported in Ref.~\cite{Han2024}. The competition between TPD strength $|\varepsilon_2|$ and ETPL rate $\kappa_2$ directly tunes the position of the LEP3 in parameter space.

\subsection{Phase-controllable Liouvillian exceptional structures}

So far, we have focused on the real-$\alpha$ case (i.e., $\theta = 3\pi/2$), which exhibits well-defined LEP2 lines and LEP3s. We now show that these Liouvillian exceptional structures can be continuously tuned by varying the TPD phase $\theta$.
As shown in Fig.~\ref{Fig3}(a), we plot the normalized LEP3 coordinates ($|\varepsilon_{|\alpha_{0}|,\,\theta}|/|\varepsilon_{|\alpha_{0}|,\,\theta_{0}}|$, $|\Delta_{|\alpha_{0}|,\,\theta}|/|\Delta_{|\alpha_{0}|,\,\theta_{0}}|$) as functions of $\theta/\pi$, with fixed reference values of $\theta_{0}=3\pi/2$ and $|\alpha_{0}|\approx1.36$.
The curves exhibit a clear periodic dependence with period $2\pi$.
Specifically, the normalized drive strength $|\varepsilon_{|\alpha_{0}|,\,\theta}|/|\varepsilon_{|\alpha_{0}|,\,\theta_{0}}|$ increases rapidly as $\theta$ approaches $\pi/2$, diverging near this value. This divergence arises from the denominator of Eq.~\eqref{LEP3_var_theta}, where the term $\mathcal{D}_{\theta}$ tends to zero in the limit of large $|\alpha|$ (i.e., $p\rightarrow1$) when $\theta = \pi/2$. Consequently, the LEP3 ceases to exist at this phase. For $\theta > \pi/2$, the curve decreases symmetrically, returning to unity at $\theta = 3\pi/2$, and completing one full period at $\theta = 2\pi$.
In contrast, the normalized detuning $|\Delta_{|\alpha_{0}|,\,\theta}|/|\Delta_{|\alpha_{0}|,\,\theta_{0}}|$ remains nearly constant across $\theta$, except for a sharp dip near $\theta = \pi/2$ where it vanishes due to the same denominator singularity. These behaviors are fully consistent with the analytic expressions in Eqs.~\eqref{LEP3_var_theta} and \eqref{LEP3_Delta_theta}.

To further illustrate the phase-dependent topology, we plot the contours of $\mathcal{R}_{1}=0$ and $\mathcal{R}_{2}=0$ for $\theta=0$ [in Fig.~\ref{Fig3}(b)], $\theta=\pi/2$ [Fig.~\ref{Fig3}(c)], and $\theta=5\pi/4$ [Fig.~\ref{Fig3}(d)], respectively.
For $\theta = 0$ and $\theta = 5\pi/4$, the contours intersect at distinct points---corresponding to LEP3s---located at approximately $(\pm 1.41, \pm 1)$ and $(\pm 1.08, \pm 1)$ in the normalized $(\varepsilon,\,\Delta)$ plane, respectively. However, at $\theta = \pi/2$, the $\mathcal{R}_1 = 0$ and $\mathcal{R}_2 = 0$ curves no longer intersect, confirming the absence of an LEP3 due to the aforementioned divergence.
Moreover, the shape and position of the LEP2 lines ($\mathcal{R}_1 = 0$) are also visibly modified by $\theta$, reflecting the phase-sensitive coupling between the logical subspace and the SPL channel.
These results establish that both LEP2 lines and LEP3s in the ETPL cat-qubit model can be coherently controlled via the TPD phase $\theta$, providing a tunable parameter for engineering NH topological features in dissipatively stabilized quantum systems.

\subsection{Validation of the Liouvillian dynamics}

\begin{figure}[b]
	\includegraphics[width =0.32\textwidth]{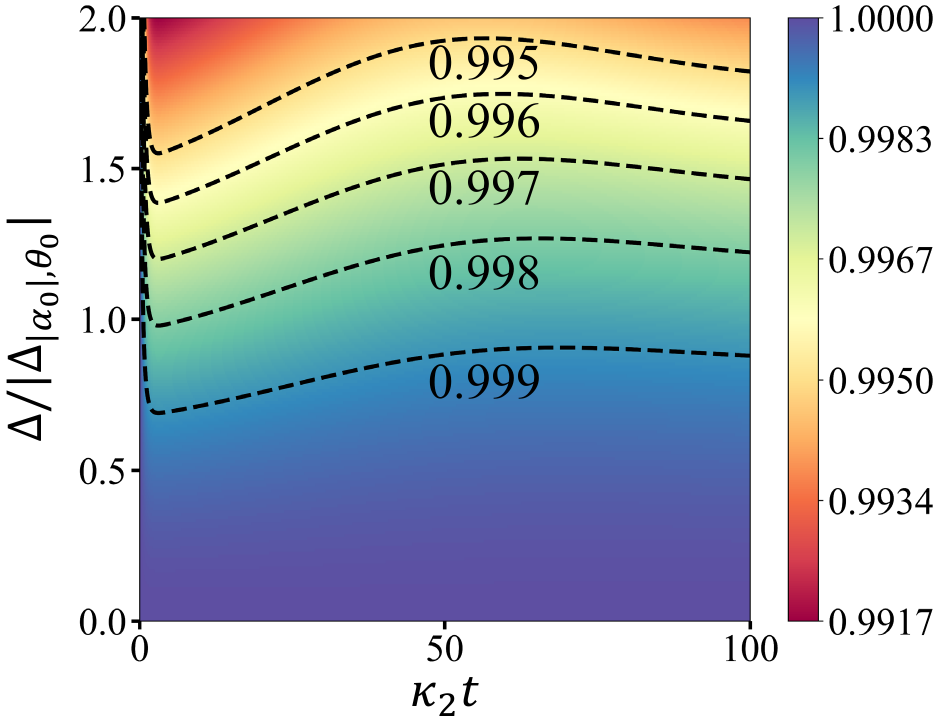}
	\caption{\label{Fig4} Validation of the Liouvillian dynamics.
    Fidelity $F(\rho_{\mathrm{H}}, \rho_{\mathrm{L}})$ between the full Lindblad evolution ($\rho_{\mathrm{H}}$) and the projected Liouvillian dynamics ($\rho_{\mathrm{L}}$) as a function of $\kappa_{2}t$ and normalized detuning $\Delta / |\Delta_{|\alpha_0|, \theta_0}|$.
    The initial state is $|\mathcal{C}^{+}_{\alpha}\rangle$, and the parameters are $\varepsilon/\kappa_{2} = 6.94\times10^{-3}$, $\kappa/\kappa_{2}\approx6.48\times10^{-3}$, $\theta_{0}=3\pi/2$, and $|\alpha_{0}|\approx1.36$.
	}
\end{figure}

To assess how the system dynamics remains confined within the ETPL cat-qubit subspace, we numerically simulate the full master equation~\eqref{eq:master}, yielding the exact density matrix $\rho_{\mathrm{H}}$, and compare it with the reduced dynamics obtained from the projected Liouvillian in Eq.~\eqref{eq:V(t)}, denoted by $\rho_{\mathrm{L}}$.
The validity of the approximation is quantified by the fidelity $F(\rho_{\text{H}},\rho_{\text{L}})=(\text{Tr}\sqrt{ \sqrt{\rho_{\text{H}}} \rho_{\text{L}} \sqrt{\rho_{\text{H}}} })^{2}$.
Fig.~\ref{Fig4} shows $F(\rho_{\text{H}},\rho_{\text{L}})$ as a function of $\kappa_{2}t$ and $\Delta/|\Delta_{|\alpha_{0}|,\,\theta_{0}}|$ in the presence of SPD. The initial state is chosen as the even cat state $|\mathcal{C}^+_\alpha\rangle$.
Across the parameter range of interest, the projected dynamics closely tracks the full dynamics, with fidelity remaining above 0.9917 for the parameters chosen in this work. Note that all these parameters are available according to the previous works reported in Refs.~\cite{Marquet2024,Marquet2024a}, demonstrating that the subspace approximation is highly robust against detuning.
Specifically, the ETPL rate $\kappa_2/2\pi = 2.16\,\text{MHz}$ and the SPL rate $\kappa/2\pi = 14\,\text{kHz}$ yield the ratio $\kappa/\kappa_2 \approx 6.48\times10^{-3}$. 
Setting the TPD strength to $|\varepsilon_2|/2\pi = 2\,\text{MHz}$ and the SPD to $\varepsilon/2\pi = 15\,\text{kHz}$ results in a normalized drive strength $|\varepsilon_2|/\kappa_2 \approx 0.93$ and a steady-state coherent amplitude $|\alpha_0| \approx 1.36$.
Moreover, the fidelity asymptotically approaches unity as the system relaxes to its steady state at long times.
The minor infidelity observed for increasing detuning arises as such a detuning term $\Delta a^\dagger a$ is only approximately captured in the effective Liouvillian description.
The detuning lifts the degeneracy between the two cat states, resulting in leakage out of the logical subspace.
Though this leakage occurs, it can be compensated by the ETPL, which effectively confines the dynamics to the cat-qubit manifold.

\section{Conclusions}
In summary, we have shown that the ETPL cat-qubit platform supports rich Liouvillian exceptional structures whose topology can be coherently controlled by the TPD phase. By constructing a resultant-vector-based topological invariant, we identified second- and third-order LEP2s and LEP3s in the parameter space spanned by SPD strength and detuning.
We further showed that the LEP3 position is highly sensitive to the TPD phase $\theta$: it diverges and vanishes at $\theta = \pi/2$, while remaining stable and tunable for other phases. This phase controllability extends to the entire LEP2 manifold, enabling on-demand reconfiguration of the NH spectral topology through a purely classical parameter.
Finally, through full master-equation simulations, we validated that the system dynamics remains tightly confined within the logical cat-qubit subspace, with state fidelity close to unity under the parameters chosen in this work. This confirms the robustness of the subspace approximation and establishes the ETPL cat-qubit as a promising platform for exploring and harnessing topological features of open quantum systems.
Our work thus establishes a direct link between dissipative stabilization, phase-coherent control, and NH topology, demonstrating that open quantum systems can serve as programmable platforms for engineering exotic spectral topologies.
This paves the way for experimental realization of higher-order exceptional points and their applications in quantum information processing.

Our results open several directions for future investigation. The phase-controlled LEPs predicted here are directly accessible in superconducting circuit implementations of dissipatively stabilized cat qubits, where the TPD phase can be tuned with high precision. Experimentally probing the winding number and the vanishing of the LEP3 at $\theta = \pi/2$ would provide a clear signature of the underlying NH topology. Beyond static spectral properties, dynamically encircling these controllable LEPs may yield chiral state transfer and non‑reciprocal responses that are robust against dissipation. Furthermore, the topological invariant introduced here can be generalized to other CV open systems, potentially enabling the design of noise‑protected quantum gates and fault‑tolerant operations that harness higher‑order exceptional physics.

\begin{acknowledgments}
This work was supported by the National Natural Science Foundation of China (Grant Nos. 12475015, 12274080, 12474356, 12505021, and 11875108), the Quantum Science and Technology-National Science and Technology Major Project (Grant No. 2021ZD0300200), and the Natural Science Foundation of Fujian Province (Grant No. 2025J01383).
\end{acknowledgments}


\nocite{*}

\bibliography{Ref}

\end{document}